\documentclass[12pt, letterpaper]{article}

\usepackage{arxiv}

\usepackage[utf8]{inputenc} 
\usepackage[T1]{fontenc}    
\usepackage{hyperref}       
\usepackage{url}            
\usepackage{amsfonts}       
\usepackage{authblk}
\usepackage[english]{babel}
\usepackage{filecontents}
\usepackage{wrapfig}
\usepackage{bm}
\usepackage{amsmath}
\newcommand{\STM}{{\mathchoice{}{}{\scriptscriptstyle}{} STM}}
\newcommand{\PC}{{\mathchoice{}{}{\scriptscriptstyle}{} PC}}
\usepackage[pdftex]{graphicx}
\DeclareGraphicsExtensions{.pdf,.jpeg,.png}

\usepackage[backend=bibtex,style=numeric-comp,sorting=none]{biblatex}
\bibliography{RC_Refs_2_arxiv.bib}

\title{Implementing a magnonic time-delay reservoir computer model}

\author{Stuart Watt}
\author{Mikhail Kostylev\thanks{Corresponding author: mikhail.kostylev@uwa.edu.au}}
\affil{Department of Physics, University of Western Australia, Crawley, W.A. 6009, Australia}

\author{Alexey B. Ustinov}
\author{Boris A. Kalinikos}
\affil{Department of Physical Electronics and Technology, St. Petersburg Electrotechnical University, St. Petersburg 197376, Russia}

\begin{document}
\maketitle

\begin{abstract}
Recently we demonstrated experimentally that microwave oscillators based on the time delay feedback provided by traveling spin waves could operate as reservoir computers. In the present paper, we extend this concept by adding the feature of time multiplexing made available by the large propagation times/distances of traveling spin waves. The system utilizes the nonlinear behavior of propagating magnetostatic surface spin waves in a yttrium-iron garnet thin film and the time delay inherent in the active ring configuration to process time dependent data streams. Higher reservoir dimensionality is obtained through the time-multiplexing method, emulating ‘virtual’ neurons as temporally separated spin-wave pulses circulating in the active ring below the auto-oscillation threshold. To demonstrate the efficacy of the concept, the active ring reservoir computer is evaluated on the short-term memory and parity check benchmark tasks, and the physical system parameters are tuned to optimize performance. By incorporating a reference line to mix the input signal directly onto the active ring output, both the amplitude and phase nonlinearity of the spin waves can be exploited, resulting in significant improvement on the nonlinear parity check task. We also find that the fading memory capacity of the system can be easily tuned by controlling the active ring gain. Finally, we show that the addition of a second spin-wave delay line configured to transmit backward volume spin waves can partly compensate dispersive pulse broadening and enhance the fading memory capacity of the active ring. \par
\end{abstract}

\keywords{Reservoir computing \and active-ring resonator \and spin waves \and magnetic thinfilm}

\section{Introduction} \label{SectionI}

A reservoir computer (RC) is a brain-inspired machine learning model which aims to simplify the construction and training of recurrent neural networks by outsourcing learning to a readout layer only \cite{Jaeger2004, Verstraeten2007}. A RC is characterized by the ‘reservoir’, which is a collection of randomly interconnected nonlinear nodes with input and output connections. The purpose of the reservoir is to apply a nonlinear kernel mapping of some low dimensional input data to a higher dimensional output space, where the input data becomes linearly separable. Recurrent connections also allow the RC to account for temporal dependencies present in the input. The output is reconstructed from the reservoir states by a linear readout, making training of a RC very simple. \par

Despite the simplicity, RCs have shown to be very powerful tools in temporal data analysis. Learning is outsourced to the readout layer, making the design of the reservoir arbitrary and not limited to simulated neurons on a computer. Any dynamical system which satisfies certain properties can be implemented as the reservoir. The first is a nonlinear mapping of the input to a higher dimensional state space to induce linear separability. The second is a ‘fading’ memory where the system state in response to the current input depends also on the recent history of inputs, whose influence over the reservoir state decays with time. \par

These ideas lead to the utilization of physical systems to implement the RC model, where the complex, nonlinear dynamics already present in nature can be substituted for simulated dynamics. The field of physical RC is fast growing and covers many disciplines due to the vast range of nonlinear dynamical systems available in nature. Physical RC offers a possible solution to obtaining fast and more energy efficient computing. Over the past several years, many different physical systems have been proposed and successfully demonstrated in order to implement the physical RC model (see e.g. the review papers by Tanaka et al. \cite{Tanaka2019} and Nakajima \cite{Nakajima2020} and references therein). \par

Among the different implementations, spintronic-based architectures are promising candidates due to the high nonlinearity of magnetization dynamics, low power usage, scalability and compatibility with existing computing technology. Many recent studies have proposed spintronic based implementations including RC architectures based on individual or arrays of coupled spin-torque nano-oscillators \cite{Torrejon2017, Furuta2018, Tsunegi2018, Markovic2019, Tsunegi2019, Riou2019,  Kanao2019, Yamaguchi2020}, magnetic skyrmion memristors \cite{Jiang2019}, magnetic skyrmion fabrics \cite{Prychynenko2018, Bourianoff2018}, dipole-coupled nano-magnets \cite{Nomura2019, Nomura2020} and spin-wave interference in garnet films \cite{Nakane2018}. The majority of works on spintronic RC have been carried out theoretically or through simulations, with only a few works \cite{Torrejon2017, Tsunegi2018, Markovic2019, Tsunegi2019, Riou2019} showing experimental results. In our recent publication \cite{Watt2020}, we added another RC concept based on the spin-wave delay-line active-ring resonator and experimentally demonstrated adequate performance on some benchmark tasks. This system utilizes the delay and nonlinear behavior of traveling spin waves in magnetic film feedback rings, and naturally satisfies the required properties of a suitable reservoir implementation. \par

In the present work, we propose a RC implementation using the method of time-multiplexing, which boosts the dimensionality of the active-ring reservoir by encoding the inputs across a series of pulses, representing ‘virtual’ nodes separated temporally along the active ring delay line \cite{Appeltant2011}. In contrast to our previous work, the proposed RC operates below the auto-oscillation threshold and exhibits much better use of the time delay of the spin-wave delay line by implementing the time-delay RC model. Similar work on physical RC using Spintronic devices with time-delayed feedback has been carried out in Refs. \cite{Riou2019, Yamaguchi2020}. Here the addition of an external feedback loop to a spin-torque nano-oscillator was shown to improve the memory capacity of the reservoir. In this work we demonstrate that the intrinsic delay time of traveling spin waves can be used to implement the same model without the need for external delay.\par

The structure of this paper is as follows. In Section \ref{SubSectionIIa} we introduce the experimental system and describe the physical dynamics behind spin-wave delay lines and the active ring construction. The RC model is described mathematically in Section \ref{SubSectionIIb} and we demonstrate how the spin-wave delay-line active-ring resonator can be utilized to implement the time delay-based reservoir. Section \ref{SubSectionIIc} describes the two benchmark tasks we employ to evaluate the performance of this system as a RC implementation. Results are presented in Section \ref{SectionIII} where the STM and PC performance is measured compared to several parameters which characterize the experimental setup. In Section \ref{SectionIV} we present a modified version of our RC where a second spin-wave delay line is added to the active ring, which yields further improvement of performance on the benchmark tasks. \par

\section{Methods} \label{SectionII}

\subsection{Description of experimental setup} \label{SubSectionIIa}

\begin{figure*}[!t]
	\centering
	\includegraphics[width=6in]{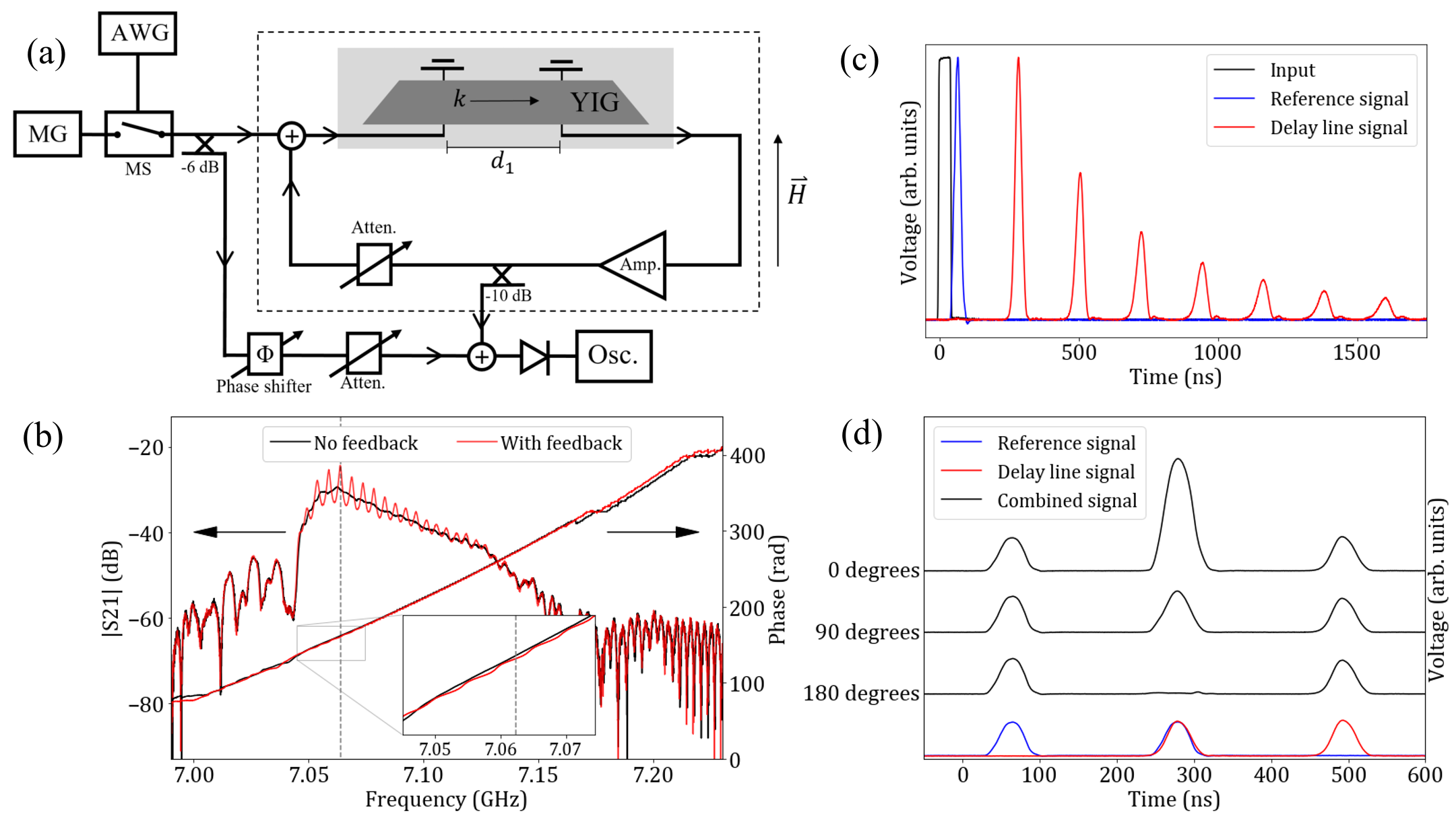}
	\hfil
	\caption{(a) Schematic diagram of the spin-wave delay-line active-ring resonator system. Components are described in the text. (b) The amplitude and phase of the transmission (S$_{12}$ parameter) characteristic for the MSSW delay line, measured with a vector network analyzer. The red trace shows resonance mode formation when feedback is added. (c) Time evolution of a single pulse injected into the active ring. The 45 ns voltage pulse from the AWG (black) creates a microwave pulse, which travels through the reference line (blue) and through the active ring (red). The traces have been normalized. Time separation between pulses is 215 ns. (d) Double pulse interference of reference (blue) and delay line signals (red) for various reference signal phases.}
	\label{fig:Fig1}
\end{figure*}

The experimental setup is shown schematically in Fig. \ref{fig:Fig1}(a). The active ring (inside the dashed box) consists of a spin-wave delay line (shaded rectangle) with a feedback loop. The delay line is assembled from a 2 mm wide, 5.7 $\mu$m thick yttrium-iron garnet (YIG) film (dark-gray trapezoid), sitting on top of two 50 $\mu$m wide short-circuited microstrip transducers (‘spin-wave antennas’). These antennas are spaced 8.2 mm apart and are used to excite/detect spin waves. \par

A spin wave is a fundamental type of magnetization dynamic in magnetic materials, representing collective precession of elementary magnetic moments (spins of localized electrons), whereby spins at neighboring crystal lattice sites are coupled via magnetic exchange and dipole-dipole interactions and the precession phase varies harmonically in one direction in space. The spin-wave frequency depends on the strength of the static magnetic field applied to the film, the material's magnetic parameters (saturation magnetization, magnetic anisotropies and gyromagnetic ratio) and also on the sample's geometry with respect to the magnetic field. In this work, the static magnetic field is applied in the plane of the YIG strip and parallel to the spin-wave antennas. In this configuration, the spin waves are called magnetostatic surface spin waves (MSSW) and travel perpendicular to the antennas. The angular frequency of the MSSW is determined by the dispersion relation \cite{Prabhakar2009} \par

\begin{equation}
\omega(H, k) = \gamma\sqrt{H(H+4 \pi M_{s})+\frac{(4 \pi M_{s})^{2}}{4}(1-e^{-2 k L})}.
\label{eq:1}
\end{equation}

Here $H$ is the external magnetic field, applied perpendicular to the spin wave propagation direction, and $k$ is the spin-wave wavenumber. The constants $|\gamma|/2\pi=2.8$ MHz/Oe, $4 \pi M_{s} = 1750$ G and $L = 5.7$ $\mu$m are the gyromagnetic ratio, the saturation magnetization and the film thickness respectively. \par

The spin-wave delay line works as follows. A microwave current injected into the input antenna (left hand side) induces a microwave Oersted field, which drives precession of the magnetic moments localized near the antenna. Neighboring moments will precess due to exchange and dipole-dipole interactions, and precession is carried away from the antennas. Energy is carried along the YIG strip in the form of a spin wave. The inverse process occurs where the dynamic magnetic dipole field produced by the precessing moments induces an electro-motive force in the output antenna (right hand side). \par  

The choice of MSSW has three benefits. First, MSSW are efficiently excited by spin-wave antennas. Second, the MSSW are unidirectional and will only be excited efficiently in one direction. This prevents energy lost to spin waves excited in the opposite direction. Finally, MSSW have a large free propagation length in epitaxially grown mono-crystalline YIG films due to very low microwave losses allowing long propagation times. \par 

Placing the spin-wave delay line into the microwave signal path introduces a time delay and phase shift of the signal determined by the spin-wave group velocity and the spin-wave antenna separation. Amplifying the output signal and feeding back into the input antenna creates an active-ring resonator \cite{Castera1978, Ishak1983, Fetisov1998} with resonant frequencies satisfying the condition $k_{res}d=2 \pi m$. Here $k_{res}$ is the spin-wave wavenumber, $d$ the antenna separation and $m$ an integer. We assume that the phase shift associated with the electrical components in the ring is negligible. \par 

Spin waves in YIG films are highly nonlinear. For the frequencies employed in this work, the four-wave nonlinear processes determine the wave nonlinearity. The processes manifest themselves as nonlinear spin-wave damping \cite{Scott2004} and a nonlinear shift of frequency or accumulated phase, depending on the experimental conditions \cite{Ustinov2006, Ustinov2008}. Because of this, YIG film delay-line active rings have been extensively studied for the host of nonlinear behavior they exhibit such as soliton formation, modulational instability, and chaos \cite{Wu2010}. We recently demonstrated how these active rings can effectively be constructed to implement the physical RC model \cite{Watt2020}. That work exploited another useful feature of active rings called auto-oscillation. At low feedback gain, delay line losses dominate and the spin-waves do not have enough power to propagate between the antennas. However, above a threshold when the feedback gain is sufficient to compensate the delay line losses, thermally excited magnons with wavenumbers satisfying the above resonance condition will be resonantly amplified. We define the ring gain $G$ as the difference between the delay line losses and the feedback gain. Thus, at the auto-oscillation threshold, $G=0$. \par

The work in Ref. \cite{Watt2020} made use of the nonlinear damping of spin waves and the slow transient response of the active ring signal to changes in $G$. In the present work, a similar active ring structure is used but operated below the auto-oscillation threshold so that the input data, injected using an external microwave signal fixed to a resonance frequency of the ring, will gradually lose power with each circulation. This gradual decay of signal amplitude implements the fading memory. The operation of the active ring below the auto-oscillation threshold is one of the key differences between this work and that of Ref. \cite{Watt2020}, providing a new physics underlying the RC operation. \par

A microwave generator (‘MG’) produces a constant signal, which is converted into a series of pulses using a PIN diode microwave switch (‘MS’) controlled using an arbitrary waveform generator (‘AWG’). The data to be processed is encoded onto this input by varying the pulse amplitude. The signal is fed into the active ring using a microwave combiner (plus symbol in Fig. \ref{fig:Fig1}(a)) after part of the signal is first sampled using a directional coupler (-6 dB) to act as a reference. The microwave pulses excite MSSW pulses at the input antenna, which travel along the YIG strip and convert back to microwave pulses at the output antenna. The signal is amplified with a low noise amplifier (‘Amp.’) and fed back into the input antenna. Since the microwave carrier frequency matches a resonance frequency of the active ring, the signal will interfere constructively with each circulation in the ring. \par

The active ring signal is sampled with another directional coupler (-10 dB), placed after the amplifier, and combined with the reference signal. A tunnel diode rectifies the combined microwave signal. An oscilloscope (‘Osc.’) measures the resulting voltage. A phase shifter in the reference line adjusts the relative phase between the signal from the active ring and the reference, and a variable attenuator adjusts the reference amplitude. Finally, a second variable attenuator in the active ring feedback loop controls the amount of feedback amplification in the ring. \par

\begin{figure*}[!t]
	\centering
	\includegraphics[width=6in]{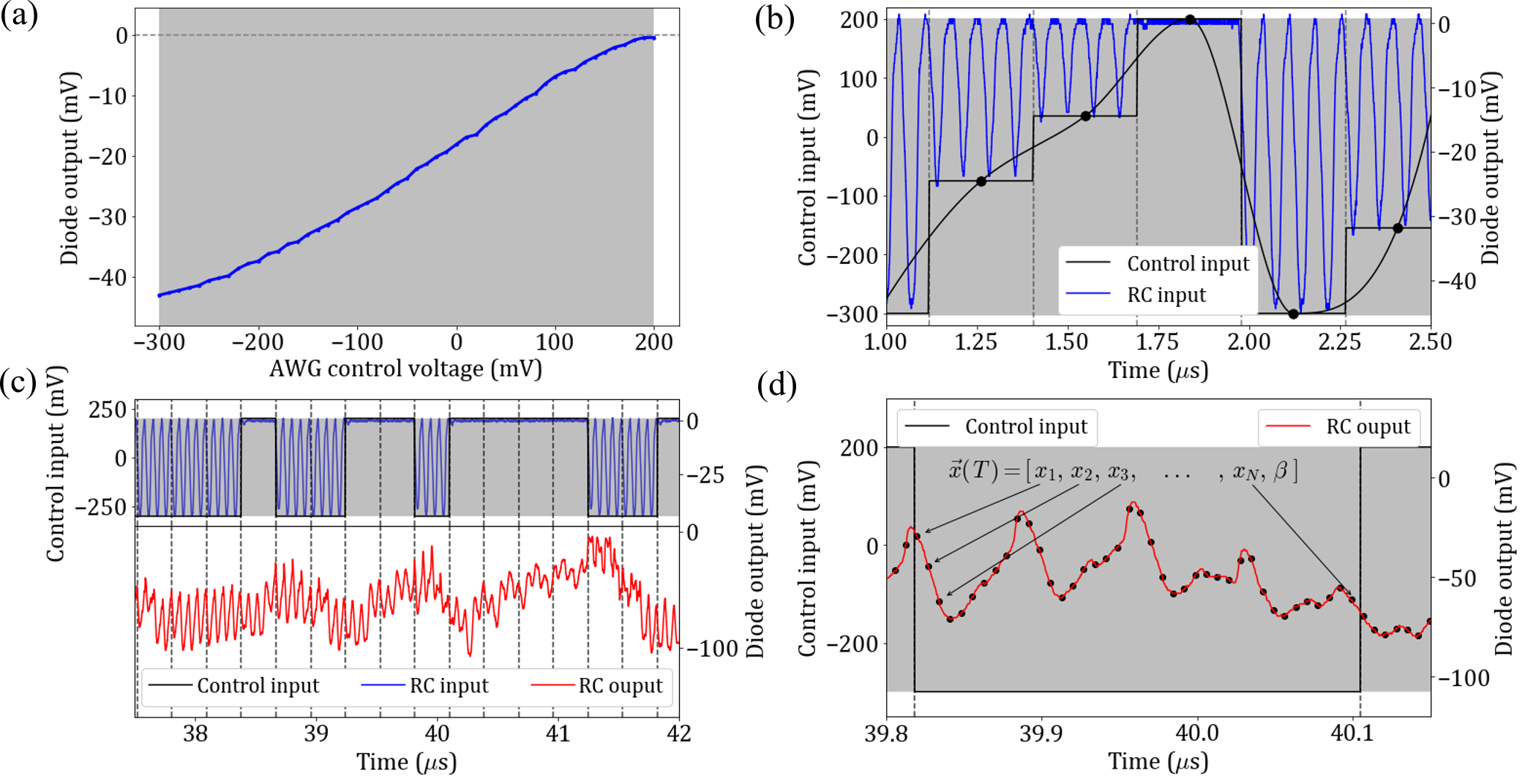}
	\hfil
	\caption{(a) Pulse amplitude from the microwave switch as a function of the AWG input control voltage (note that the diode is inverting). The gray region indicates the computing range of input control voltage. (b) Example of a continuous signal sampled and encoded using $n=4$ pulses. The ‘control input’ represents the amplitude of the voltage pulses from the AWG, and the ‘RC input’ shows the microwave pulses created by the MS and measured with the microwave diode. (c) Example of binary input (blue) and output (red) for the STM and PC tasks. (d) Example of sampling each output into $N$ virtual nodes to create the system state vector $\vec{x}(T)$. To each output a constant $\beta=0.1$ is added, making the output dimension $N+1$.}
	\label{fig:Fig2}
\end{figure*}

A static magnetic field of 1772 Oe is produced by two permanent neodymium magnets on a cast iron yoke. Fig. \ref{fig:Fig1}(b) shows the amplitude-frequency and phase-frequency characteristics of the delay line. When feedback is added to the delay line, ring resonance modes are formed. As the ring gain is raised above the auto-oscillation threshold, the resonance mode with the lowest loss is self-excited first, followed by the second lowest loss resonance mode and so on. In this case, the frequency of the resonance mode with the lowest loss is 7.06 GHz and the external microwave signal is set to this frequency. As mentioned above, the ring is operated \textit{below} the auto-oscillation threshold and so all inputs will gradually fade over time. To demonstrate this, Fig. \ref{fig:Fig1}(c) shows the time evolution of a 45 ns input pulse as it circulates in the active ring below the auto-oscillation threshold. The reference signal (blue) is delayed by about 40 ns from the AWG control voltage. After injection into the ring (red), the pulse circulates many times, its amplitude decreasing with each successive cycle. YIG is a dispersive medium, which results in dispersive pulse broadening. The pulse energy spreads out over time, reducing the peak amplitude. The dispersion coefficient is determined as \par

\begin{equation}
D=\frac{\partial^{2} \omega(H, k)}{\partial k^{2}}.
\label{eq:2}
\end{equation}

The time delay between the reference pulse and the pulse after one period of circulation in the ring is 215 ns, which matches the separation between successive circulations of the pulse. This indicates that the time delay in the feedback loop is negligible and that the circulation time is due predominantly to the spin-wave delay line. \par

\subsection{RC operation} \label{SubSectionIIb}

The RC model has received much attention due to its simple structure of just three components \cite{Lukosevicius2012} - the input, reservoir and read-out. When constructed using software, the reservoir comprises a series of nodes with weighted connections between the input layer, output layer, and internal reservoir nodes. The reservoir state at some time step, $T$, is described by a vector, $\vec{x}(T)$, containing the values of each reservoir node and is updated as \par

\begin{equation}
\vec{x}(T)=f( \bm{W^{in}} \vec{u}(T) + \bm{W} \vec{x}(T-1)).
\label{eq:3}
\end{equation}

The matrices $\bm{W^{in}}$ and $\bm{W}$ determine the input and internal connection weights respectively and $\vec{u}(T)$ is the vector of input values. The reservoir dimensionality is usually much higher than that of the input, with a large number of nodes characterized by a nonlinear activation function, $f$. The purpose of the reservoir then is to perform a kernel mapping from lower to higher dimensions. Unlike the recurrent neural networks, where all the connection weights are trained using the back-propagation through time method, $\bm{W^{in}}$ and $\bm{W}$ are randomly initialized and remain fixed throughout the training. Only the readout weights are trained. The output of the system is obtained as a weighted sum (linear regression) of the reservoir states

\begin{equation}
\vec{y}_{out}(T)=\bm{W^{out}} \vec{x}(T).
\label{eq:4}
\end{equation}

During the training phase, the weights $\bm{W^{out}}$ are adjusted to reduce the mean squared error between $\vec{y}_{out}(T)$ and some target output. When implemented physically, $\bm{W}$ and $f$ are determined by the physical parameters of the implementation, while $\bm{W^{in}}$ is determined by the input mechanism to the system. Presently, the dynamics of the microwave signal in the active ring determine $\bm{W}$ and $f$, while the AWG control voltage applied to the MS determines $\bm{W^{in}}$. \par

The inputs are injected into the active ring by encoding them onto a train of microwave pulses with a width of 45 ns and a repetition period of 71.7 ns, thus allowing for three pulses to circulate simultaneously in the ring. A variable input is achieved by varying the amplitude of the pulses. For this, the control voltage is set within the range $V_{in}=[-300,200]$  mV. Fig. \ref{fig:Fig2}(a) shows the pulse amplitude as a function of the AWG control voltage. The gray region indicates the computing range of AWG control voltage, for which the pulse amplitude varies approximately linearly from maximum to zero. \par

Using the time multiplexing method, each input can be encoded across multiple consecutive pulses. Ref. \cite{Appeltant2011} showed how a single nonlinear node can be used to emulate an entire reservoir by substituting many spatially separate nodes with a series of virtual nodes separated temporally. The time-multiplexing method boosts the dimensionality of a single node reservoir. The RC architectures which implement this method are referred to as time-delay RC (TDRC) models. TDRC models have been extensively studied in optical and optoelectronic circuits \cite{Paquot2012, Brunner2018}. The main benefit of TDRC is in the simplicity of fabrication, requiring only a single nonlinear node in place of an entire array of nodes, each with their own inputs and outputs. \par

To implement this method in the active ring, each input is encoded across multiple pulses. The number of pulses used to encode each input is denoted by $n$. The total duration, $\theta^{int}$, of each input interval is then determined by the number of pulses ($\theta^{int}=71.7n$ ns). In the traditional time-multiplexing technique, each input is multiplied by a randomized ‘masking’ matrix to create the different virtual node inputs. In this system however, due to the limited capabilities of the available function generator, the node count per cycle is much lower than in fiber-optic systems. Therefore, this masking matrix is not randomized and instead all pulses in each input have the same amplitude. For example, Fig. \ref{fig:Fig2}(b) shows each input encoded using $n=4$. Unless otherwise stated, inputs are encoded using $n=4$ pulses. The effect of encoding the input using different $n$ is investigated in Section \ref{SubSectionIIIc}. \par

Each input is injected into the system, and the active ring state vector, $\vec{x}(T)$, for each input is measured using the fast microwave diode. Due to the time delay introduced by the spin-wave propagation, the active ring output signal at time $T$ is not influenced immediately by the current input, $\vec{u}(T)$, for that interval. In the case where $\theta^{int}$ is less than the active ring round trip time, there is no influence at all of $\vec{u}(T)$ on the delay line output and the reservoir state vector becomes \par

\begin{equation}
\vec{x}(T)=f( \bm{W^{in}} \vec{u}(T-1) + \bm{W} \vec{x}(T-1)).
\label{eq:5}
\end{equation}

The addition of the reference signal resolves this problem by adding the input directly to the output. The RC output equation then becomes \par

\begin{equation}
\vec{y}_{out}(T)=\bm{W^{out}} (\vec{x}(T) + \alpha \vec{u}(T)).
\label{eq:6}
\end{equation}

When constructing a RC model using software the input can be directly included into the linear regression in the post-processing stage. By feeding part of the input off using the reference line, this step is done physically, simplifying the post-processing without any degradation in processing time.  Furthermore, adding the reference signal to the RC output adds more tunable parameters to the system; notably the relative phase of the reference signal with respect to the active ring signal and the input scaling factor $\alpha$. \par

The scaling factor $\alpha$ represents an attenuation of the reference signal amplitude before combining it with the active ring output signal. The attenuation is constant in all measurements such that the pulse from the reference signal and the pulse after one pass through the delay line are equal in amplitude. Fig. \ref{fig:Fig1}(d) shows this case. With the feedback line disconnected, two pulses are injected into the system with a separation equal to the delay time (215 ns). The blue and red traces show the reference signal and delay line signal respectively, while the black traces shows the combined signal. The pulses interfere constructively or destructively depending on their relative phase. This interference provides an extra layer of complexity and nonlinearity to the diode output. \par

Fig. \ref{fig:Fig2}(c) shows an example of the system output in response to a binary input. For each input interval, the output diode voltage is sampled into $N$ equispaced points to obtain the virtual node values, as shown in Fig. \ref{fig:Fig2}(d). This creates an output vector, $\vec{x}(T)$, for each input with a dimensionality of $N$. In this work we choose to sample the data such that each pulse is separated into 10 equispaced nodes, making a total of $N=10n$ virtual node values in each output. An additional bias term, $\beta=0.1$, is concatenated to $\vec{x}(T)$ for all time intervals in the post-processing stage. This term corresponds to one additional ‘virtual’ neuron which adds a constant to the linear combination of the output. The non-zero value of $\beta$ was chosen arbitrarily and does not affect the computational performance. The total output dimensionality is then $N+1$. \par

\subsection{Benchmark performance tasks} \label{SubSectionIIc}

To evaluate the performance of the active-ring resonator system as a RC, we employ the short-term memory task (abbreviated earlier as STM) \cite{Jaeger2002} and the parity check task (abbreviated earlier as PC) \cite{Furuta2018, Bertschinger2004}. The employed procedure is the same as in our previous work \cite{Watt2020}. However, we believe that it is useful to briefly describe it here for the sake of completeness of the paper. In both tasks, the system state at a given time $T$ is measured in response to a random binary input $u(T) \in [0,1]$. \par

The STM task provides a measure of the fading memory present in the system. The target for each time step is simply the input at some delay ($\tau$ steps) in the past determined by \par

\begin{equation}
\hat{y}_{\STM}(T, \tau) = u(T - \tau).
\label{eq:7}
\end{equation}

The PC task is a non-linearly separable task, which requires both fading memory and nonlinearity. The target for each time step is determined by taking the parity of the sum of the consecutive inputs up to some delay in the past \par

\begin{equation}
\hat{y}_{\PC}(T, \tau) = PARITY[u(T-\tau), u(T-\tau+1),...,u(T)].
\label{eq:8}
\end{equation}

Here the PARITY operation returns the parity (0 for even, 1 for odd) of the sum of the values in the brackets. In both tasks, the targets are also binary values. \par

A sequence of 2200 random binary inputs is fed into the RC to perform the tasks. The inputs [0,1] are converted to corresponding AWG input control voltages [200 mV,-300 mV] (no pulse or maximum amplitude pulse) respectively. Fig. \ref{fig:Fig2}(c) shows an example of the input and output for these tasks. \par

\begin{figure}[!t]
	\centering
	\includegraphics[width=3in]{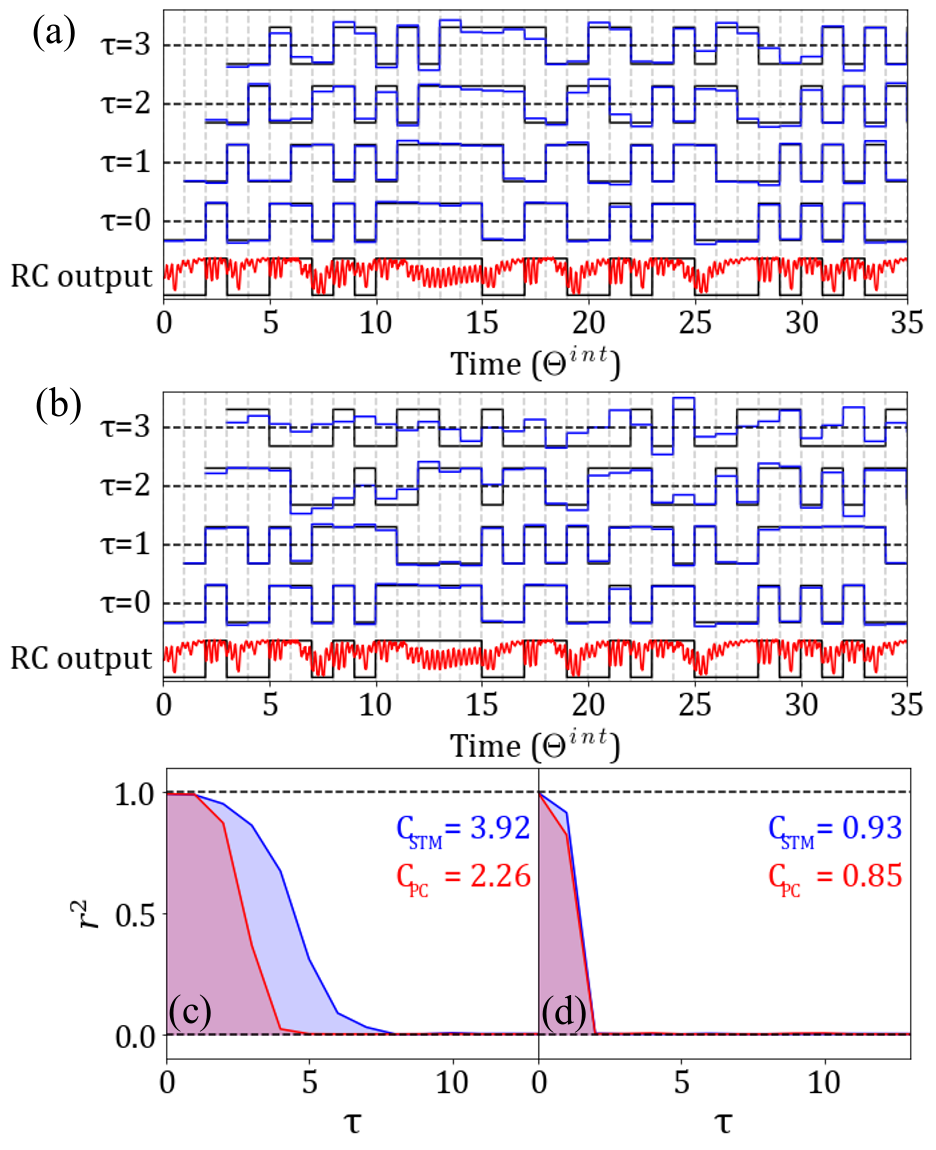}
	\hfil
	\caption{(a) Comparison of the targets, $\hat{y}_{STM}(T)$ (black), and reconstructed targets, $y_{out}$ (blue), for the linear regressions trained on the RC output (red) for the STM task. (b) Same as for (a) but for the PC task. (c) Square of the correlation coefficient for increasing delay (forgetting curve) for the STM and PC tasks. The data shown here is for a ring loss of 7.68 dB and a binary input encoded using $n=4$ pulses. The reference signal is in anti-phase to the active ring. (d) Forgetting curve for the STM and PC tasks evaluated using only the reference signal.}
	\label{fig:Fig3}
\end{figure}

Since both tasks require the same input, they can be performed simultaneously. Conventional recurrent neural networks are trained for specific tasks and are then not generalizable. In RC, only the read out weights are trained, allowing the same reservoir to be adapted to a wide range of tasks. This multitasking property is especially useful when implementing the RC physically, since the physical parameters of the system need not be altered to execute a specific task. \par

The entire 2200-value sequence is fed into the active ring and 2200 output vectors are recorded in one batch. Training of $\bm{W^{out}}$ and evaluation can all be done ‘off-line’ later. The first 200 outputs are discarded. $\bm{W^{out}}$ is trained on the following 1000 outputs and evaluated on the remaining 1000 outputs. Defining $\bm{Y}$ and $\bm{X}$ as matrices containing the targets and reservoir states for all time steps in the training set (i.e. $\bm{X}$ is the matrix of reservoir state vectors $\vec{x}(T)$ for all $T$ concatenated horizontally, similarly for $\bm{Y}$), the optimal $\bm{W^{out}}$ is obtained by taking the product $\bm{Y} \bm{X}^{-1}$. This singular training step is stable and very fast. A different $\bm{W^{out}}$ is trained for each value of $\tau$. The success of the linear regression to reconstruct the desired target is measured by calculating the square of the correlation coefficient between the reconstructed target, $y_{out}$, and the actual target, $\hat{y}$ \cite{Furuta2018}. \par

Fig. \ref{fig:Fig3}(a) and (b) show examples of the comparison between the reconstructed target and the actual target for the STM and PC tasks respectively. The x-axis is denoted in units of $\theta^{int}$. The black traces show the targets, $\hat{y}$, for each input interval while the blue traces show the reconstructed (predicted) targets, $y_{out}$, after the readout weights have been trained. As $\tau$ is increased, the ability to correctly predict the target from the current output is reduced. \par

This behavior is summarized using ‘forgetting’ curves (shown in Fig. \ref{fig:Fig3}(c) and (d)) which plot $r(\tau)^{2}$ against $\tau$. These curves are visual aids to show how the reconstruction performance depends on the delay. Finally, taking the sum of $r(\tau)^{2}$ over the range of $\tau$ (equivalent to the area under the forgetting curves) returns the STM and PC capacities as  \par

\begin{equation}
C_{\STM/\PC} = \sum_{\tau=1}^{\tau_{max}=20} r(\tau)^{2}.
\label{eq:10}
\end{equation}

Ten different 2200-value binary sequences are passed through the system and the capacities are calculated for each trial and averaged. The uncertainties in $C_{\STM}$ and $C_{\PC}$ are calculated as the standard deviation across the ten trials. \par

\section{Results and Discussion} \label{SectionIII}

In this section, we explore how the STM and PC task performance depends on the various tunable parameters in the system. In order to obtain meaningful estimations of the system performance, the contribution from the active ring must be isolated from that of the pre-processing and input generation steps (nonlinearities and transient behaviors arising from the MS). The STM and PC capacities are first measured using only the signal from the reference line, essentially removing the active ring completely. The output equation becomes \par

\begin{equation}
y_{out}(T)=\bm{W^{out}} (\alpha u(T)).
\label{eq:11}
\end{equation}

Since $\bm{W^{out}}$ is linear and the reference signal is instantaneous and memory-less, the STM and PC capacities should ideally be 0. This is not the case however, as can be seen in the forgetting curve in Fig. \ref{fig:Fig3}(d). \par

The reason that the $r(\tau)^{2}$ do not drop to zero for $\tau=1$ is that the pulses are not completely separated, and the forth pulse of the previous input merges slightly into the first pulse of the current input (see Fig. \ref{fig:Fig2}(b)), hence there is a slight fading memory that only goes back one unit of delay. The forgetting curve in Fig. \ref{fig:Fig3}(d) supports this claim. The linear regression can accurately predict the current input and also adequately predict the previous input. The prediction accuracy drops to 0 for a delay of $\tau = 2$ or more. \par

These benchmark measurements allow us to evaluate the influence of the active ring alone on the computation by isolating the contribution to computation due to the pre-processing steps and input generation at the MS. In the figures that follow, these benchmark capacities are displayed as horizontal dashed lines. \par

\subsection{Performance against active ring gain} \label{SubSectionIIIa}

Tuning the active ring gain controls the fading memory of the system. The low-noise amplifier provides a constant amplification and the total gain of the ring is controlled with the variable attenuator. We thus define the ring gain in terms of the loss introduced by the attenuator relative to the auto-oscillation threshold. As ring loss is increased, the inputs decay more rapidly with each circulation of the ring. This is equivalent to a reduction in memory. \par

\begin{figure}[!t]
	\centering
	\includegraphics[width=2.8in]{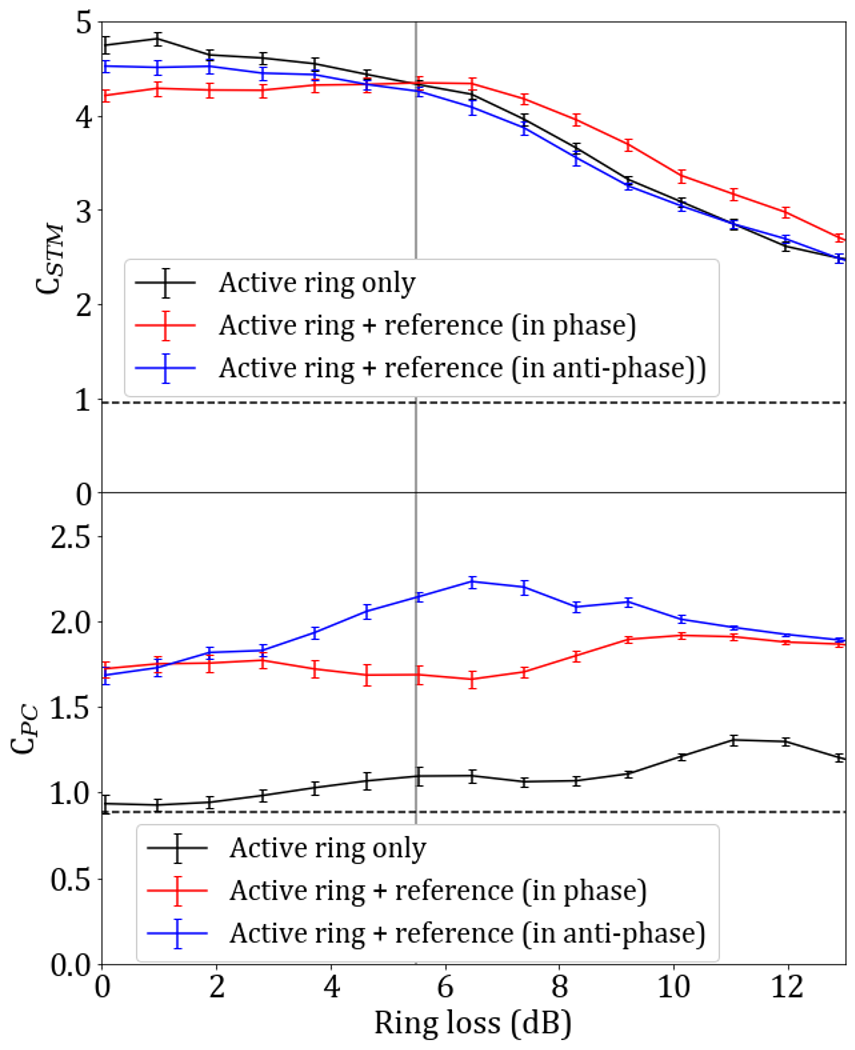}
	\hfil
	\caption{$C_{\STM}$ and $C_{\PC}$ measured for increasing loss in the active ring with $n=4$ pulse encoding. The auto-oscillation threshold is defined as 0 dB loss. The black data are measured without the reference signal. The red(blue) data are measured with the reference line connecting in phase(anti-phase) with the active ring.}
	\label{fig:Fig4}
\end{figure}

Fig. \ref{fig:Fig4} shows the STM and PC capacities measured against the ring loss. Two different operational schemes are compared. In the first (black), the reference line is disconnected. In the second, the reference line is connected and the system output depends on both the input and the active ring signals, as in Eq. \ref{eq:6}. The relative phase of the reference signal is chosen to be completely in phase (red) and in anti-phase (blue) with the active ring signal. \par

$C_{\STM}$ decreases monotonically due to the reduced pulse persistence as the ring loss is increased. Note that \textit{qualitatively} the behavior is similar between the three different cases presented. The addition of the reference signal to the output of the RC should not affect the memory capacity in the active ring because the fading memory originates in the active ring alone. There is a slight reduction in the maximum STM capacity reached when the reference signal is included, perhaps being the result of the reference signal partly dominating the active ring signal. The maximum $C_{\STM}$ reached in the first scheme is $4.82 \pm 0.0$9. For the second scheme, the maximum $C_{\STM}$ is $4.35 (4.53) \pm 0.08 (0.08)$ when the reference is in phase(anti-phase). \par

As to the PC task, the addition of the reference line significantly improves $C_{\PC}$ due to the nonlinear nature of the task, reaching a maximum of $2.23 \pm 0.05$ in the anti-phase case and only $1.31 \pm 0.05$ in the case without the reference signal. There are two reasons for this. First, the system output is not actually influenced by the current input if the reference line is disconnected, apart from the transient merging of one input into the next. The second reason is the interference of the two signals when the reference line is connected. This interference adds a degree of complexity to the output and thus helps the computation. The amplitude of an interference pattern is a nonlinear (periodic) function of the phase difference between the interfering signals adding more nonlinearity to the system. Furthermore, by combining the active ring signal with a reference one, both amplitude and phase information from the active ring signal are imparted to the output. In addition to the nonlinear damping \cite{Scott2004} exploited in Ref. \cite{Watt2020}, spin waves also exhibit a nonlinear power dependent phase shift \cite{Ustinov2006, Ustinov2008}. This additional source of nonlinearity in the reservoir aids in the quantitative improvement of the computation performance from our previous work. \par

\subsection{Performance against reference phase} \label{SubSectionIIIb}

\begin{figure}[!t]
	\centering
	\includegraphics[width=3in]{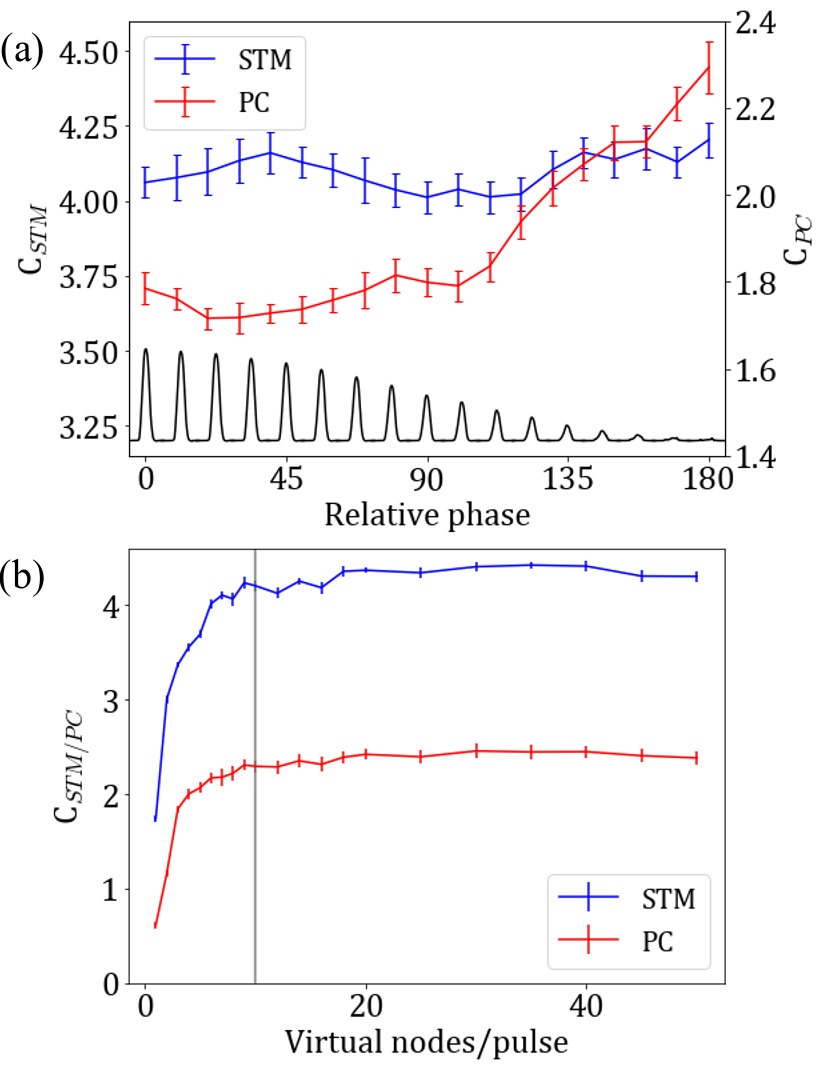}
	\hfil
	\caption{(a) $C_{\STM}$ and $C_{\PC}$ measured against the relative phase between reference and active ring signals. The black traces show the interference pattern of two pulses. (b) Dependence of $C_{\STM}$ and $C_{\PC}$ on the number of virtual nodes sampled per pulse in the output. The vertical line designates 10 nodes/pulse.}
	\label{fig:Fig5}
\end{figure}

As shown in Fig. \ref{fig:Fig4}(b), $C_{\PC}$ is significantly improved when comparing whether the reference signal is in phase or in anti-phase with the active ring. To explore this, the ring loss is set to 5.49 dB (vertical gray line in Fig. \ref{fig:Fig4}) and $C_{\STM}$ and $C_{\PC}$ are measured against the reference phase. At this point, all three cases in Fig. \ref{fig:Fig4}(a) returned the same $C_{\STM}$. Indeed this is confirmed in Fig. \ref{fig:Fig5}(a) where $C_{\STM}$ is unchanged for all reference phases. On the other hand, the PC results show a strong dependence on the reference phase, $C_{\PC}$ being maximized when the reference is in anti-phase to the active ring. \par

The traces at the bottom of Fig. \ref{fig:Fig5}(a) show the mixed pulse at the output. Note that due to dispersive broadening of the spin-wave pulses, the output pulse from the active ring is slightly wider than the pulse from the reference line, and the shoulders of the pulse remain even when the two pulses are in anti-phase. This fact may actually benefit the computational performance of the active ring RC, as the pulses get wider with each circulation of the active ring. The shape of the output trace will then become more complex. Note that for this complexity to be captured, one needs to sample a sufficient number of virtual nodes from the output. \par 

The data presented so far have used $N=10n$ virtual nodes. Fig. \ref{fig:Fig5}(b) shows how $C_{\STM}$ and $C_{\PC}$ depend on the number of virtual nodes sampled for each pulse. After approximately 10 nodes/pulse, the performance saturates. With the addition of more nodes, the nodal separation becomes shorter than the timescale of variations in the output and these extra nodes do not add any information to the linear readout. \par

\subsection{Performance against $n$} \label{SubSectionIIIc}

Here $C_{\STM}$ and $C_{\PC}$ are measured against the active ring loss when the input sequence is encoded using $n=$ 1 to 4 pulses. For these measurements, the reference signal is kept in anti-phase with the active ring signal. \par

\begin{figure}[!t]
	\centering
	\includegraphics[width=2.8in]{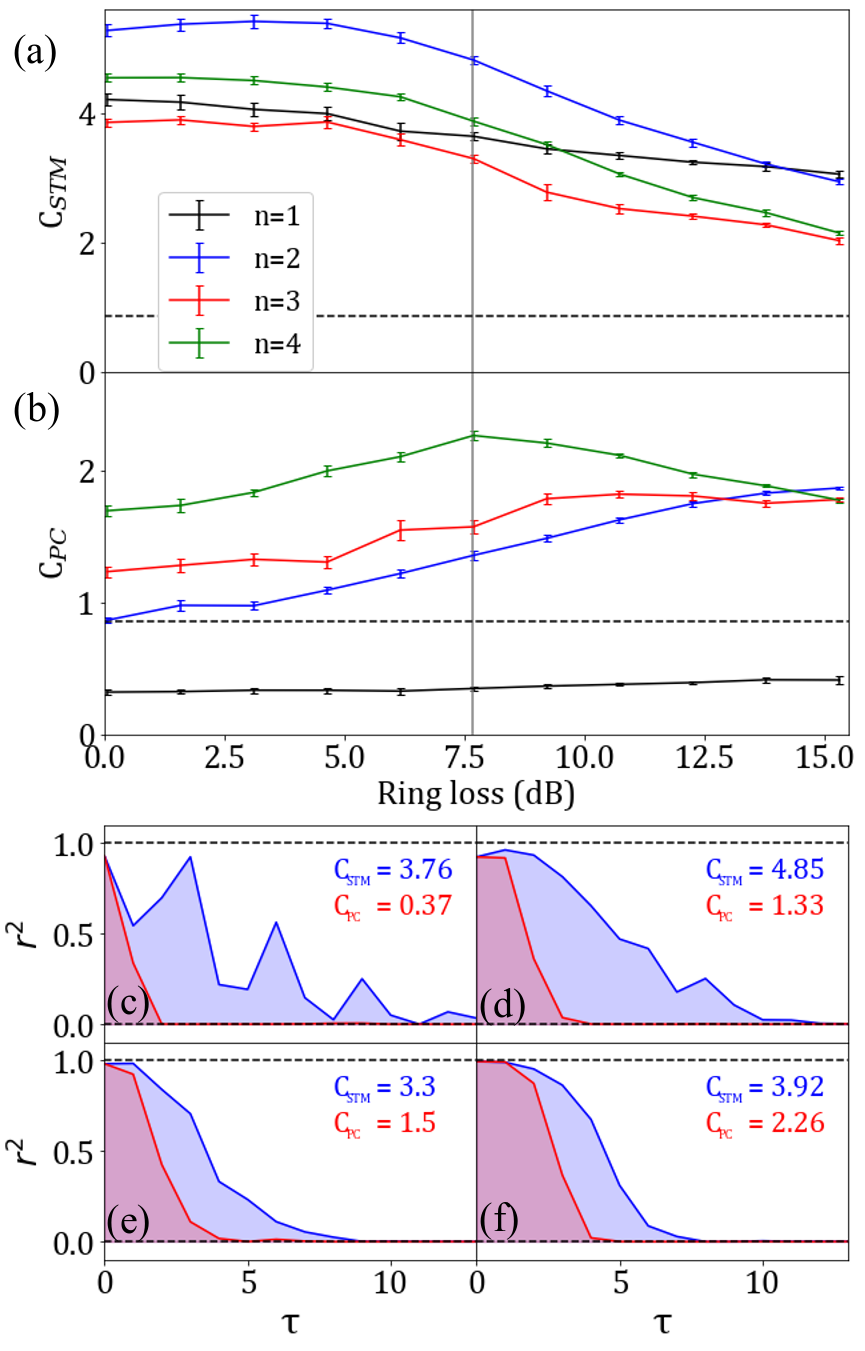}
	\hfil
	\caption{(a)-(b) $C_{\STM}$ and $C_{\PC}$ measured for increasing ring loss with $n=1$ to $4$. The auto-oscillation threshold is defined as 0 dB loss. (c)-(f) Forgetting curves corresponding to a ring loss of 7.68 dB (vertical gray line in (a)-(b)) for the $n=1$ to $4$ pulse encodings.}
	\label{fig:Fig6}
\end{figure}

Plots of $C_{\STM}$ against the feedback loss are shown in Fig. \ref{fig:Fig6}(a). For each encoding, the qualitative behavior is the same as that described in Section \ref{SubSectionIIIa}, decreasing monotonically as the fading memory is reduced. The same can be said for $C_{\PC}$ shown in Fig. \ref{fig:Fig6}(b). To help elucidate the behavior shown in the plots, the forgetting curves for each encoding are shown in Fig. \ref{fig:Fig6}(c)-(f). \par

Given that 3 pulses circulate simultaneously in one cycle of the active ring, when the input is encoded using $n=1$, there is only mixing of every third input. This is evident in the forgetting curve in Fig. \ref{fig:Fig6}(c). The short term memory recall peaks when $\tau$ is a multiple of 3 and is poor for other delays. On the other hand, the shorter input duration means that inputs are injected to the ring much faster and will persist in the ring for more input intervals before fading out. \par

The lack of mixing between consecutive inputs significantly hinders performance on the PC task. Recall that the PC task involves a binary operation over all the inputs up to some $\tau$ in the past, and so the lack of mixing between consecutive inputs and the lower number of virtual nodes has drastic effects on $C_{\PC}$. \par

\begin{figure*}[!t]
	\centering
	\includegraphics[width=6in]{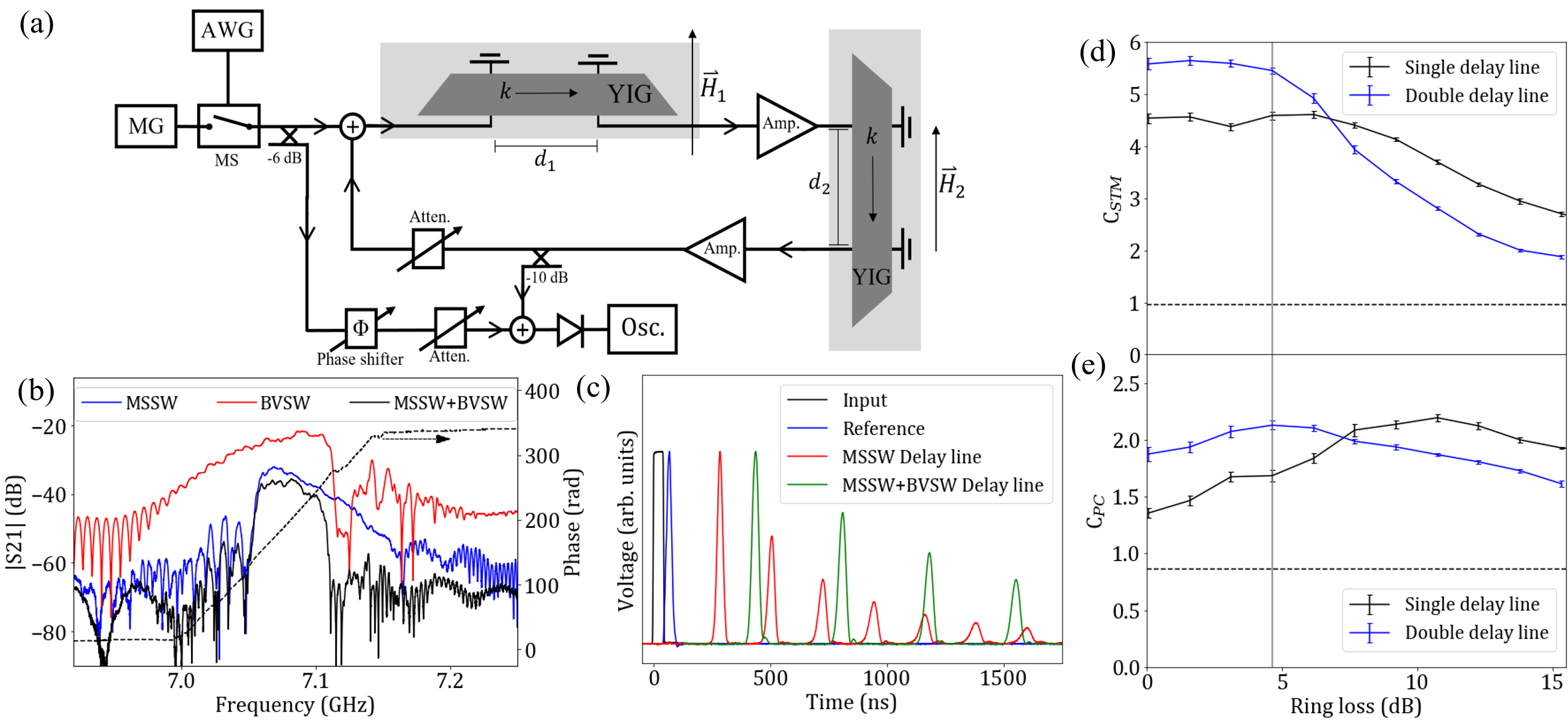}
	\hfil
	\caption{(a) Schematic diagram of the extended spin-wave delay-line active-ring resonator system. (b) Amplitude transmission characteristics for the MSSW (blue), BVSW (red) and combined (black) delay lines. Dashed black line shows the transmission phase of the combined delay line. (c) Time evolution of a single pulse injected into the active ring. The 45 ns voltage pulse from the AWG (black) creates a microwave pulse, which travels through the reference line (blue) and through the single (red) and double (green) delay-line active rings. The traces have been normalized. (d) $C_{\STM}$ measured for increasing ring loss. Data compares the single (black) and double (blue) delay-line active rings. (e) Same as (d) but for PC task.}
	\label{fig:Fig7}
\end{figure*}

The case is different once the input is encoded with $n=2$. Since the round trip time is 3 pulses long, each input partially overlaps with the previous input as well as the reference signal. This prevents any periodicity in the forgetting curve and $C_{\STM}$ reaches its maximum values. Performance on the PC task is also significantly improved due to this overlap as well as double the number of virtual nodes available for computing. However, because of the partial overlap, only half of the output nodes are influenced by the preceding input. \par

For $n=3$, there is a complete overlap between consecutive inputs. The PC and STM forgetting curves (Fig. \ref{fig:Fig6}(e)) reach almost 1 for $\tau = 0$. Combining this overlap with the larger number of virtual nodes again improves $C_{\PC}$ from the $n=2$ case. On the other hand, $C_{\STM}$ is reduced as the inputs are injected slower relative to the active ring circulation time and thus their influence does not persist for as many input intervals. Additionally, it has been shown that the memory capacity in delay-based RCs is degraded when the delay time is an integer multiple of the input time interval \cite{Yamaguchi2020}. This is related to the interaction of different virtual neurons between time steps when the delay time and input time interval are not synchronized.\par

In the final case of $n=4$, there is a partial overlap between the fourth pulse in the reference signal (current input) and the first pulse of the delayed input through the active ring. This overlap significantly increases the complexity of the output. For this reason and the greater number of virtual nodes, $C_{\PC}$ is significantly improved. One may also expect a significant improvement in $C_{\STM}$ also due to the desynchronization of the delay time to the input, however this improvement is partially compensated by the longer input time and reduced input persistence in the ring (evidenced by the sharper drop off of the forgetting curve (Fig. \ref{fig:Fig6}(f)) compared to the previous 3 cases). \par

One can expect reductions in both $C_{\STM}$ and $C_{\PC}$ as $n$ is further increased, due to the slower injection time relative to the signal attenuation. An input encoding with n=4 provides the optimal compromise between fading memory and nonlinear computing capability.  \par

\section{Extended delay line} \label{SectionIV}

In this section, the delay time of the active ring is increased by the addition of a second delay line connected in series with the first. The experimental setup is depicted schematically in Fig. \ref{fig:Fig7}(a) where the output of the first delay line is amplified and then injected into the input of the second. The additional delay line is constructed in the same way as the first. The antenna separation is $d_{2}=4.9$ mm. The defining difference between the two delay lines is the orientation with respect to the applied magnetic field. The first delay line is orientated such that the spin-wave propagation direction is perpendicular to the magnetic field, exciting MSSW. The second delay line is oriented such that the spin-wave propagation is parallel to the in-plane magnetic field, exciting backward volume spin waves (BVSW). These spin waves are characterized by the dispersion relation \cite{Kalinikos1980, Prabhakar2009}\par

\begin{equation}
\omega(H, k) = \gamma\sqrt{H[H+4 \pi M_{s}(\frac{1-e^{-k L}}{kL})]}.
\label{eq:12}
\end{equation}

The BVSW delay line is placed in a magnetic field of 1813 Oe to align the transmission band to that of the MSSW delay line. The transmission characteristics of each individual delay line as well as that of the combined delay line are shown in Fig. \ref{fig:Fig7}(b). \par

Delay-line active rings configured using either MSSW or BVSW operate in the same manner. Thus the RC performance of an active ring configured for the transmission of BVSW would yield results qualitatively similar to the preceding sections. Where the two forms of spin wave differ is their excitation efficiency and dispersion coefficient. The excitation of BVSW is not as efficient as MSSW due to the way that the components of the dynamic magnetization vector couple to the driving magnetic field. For the same microwave power, the MSSW are excited more strongly, and at the output antenna, the inverse process is likewise more efficient. Furthermore, the excitation of BVSW is not unidirectional, so half power of the input is instantly lost to BVSW traveling away from the second antenna. Thus, from an engineering viewpoint, the use of the MSSW is favorable because smaller powers are needed for the active ring operation. Alternatively, for the same available input power, the MSSW-based active ring will operate in a significantly more nonlinear regime than the BVSW-based one. For these reasons, the MSSW were chosen as the information carrier in the preceding sections.\par

On the other hand, the choice of using a BVSW delay line instead of a second MSSW delay line is due to the opposing dispersion coefficients (Eq. \ref{eq:2}) of BVSW and MSSW. MSSW have a negative dispersion coefficient. With the addition of a second MSSW delay line, the pulses would broaden more with each circulation of the active ring and their amplitude would decay faster. Since the active ring state is measured using the amplitude of the microwave signal in the ring, the faster broadening of the pulses effectively reduces the amount of memory of the system. \par

BVSW have a positive dispersion coefficient, which can partly compensate the broadening of the pulses due to the first MSSW delay line. In this way, the total active ring circulation time can be extended, without the addition of excessive pulse broadening. Fig. \ref{fig:Fig7}(c) compares the evolution of a single pulse traveling around the single or double delay-line active ring. The BVSW delay line introduces an additional 152 ns delay. In addition to this, the pulse, which circulates in the double-delay-line active ring, decreases in amplitude at a lower rate than in the single-delay-line case. Indeed, the width of the pulse in the single-delay-line active ring grows faster than in the double-delay-line case, due to the opposite dispersion coefficients. \par

The operation of the double-delay-line active ring is functionally the same. The input is encoded onto a train of 45 ns wide pulses with a repetition period of 73.4 ns, such that 5 pulses circulate simultaneously. Using the results for optimizing the active ring RC performance determined in the previous section, $C_{\STM}$ and $C_{\PC}$ are measured against the ring loss with the following parameters. The reference line is connected in anti-phase with the active ring. The input sequence is encoded using n=6 pulses (one greater than the number of pulses which simultaneously circulate in the ring to avoid input synchronization with the active ring). \par

The double-delay-line active-ring RC shows the same qualitative performance as the single-delay-line active ring, with some differences. The first is a significant improvement of $C_{\STM}$ at a low ring loss (as shown in Fig. \ref{fig:Fig7}(d)). This is most likely due to the combination of an increased number of nodes and the slower decay rate of the pulse amplitude during each round trip of the active ring.

The PC capacities (shown in Fig. \ref{fig:Fig7}(e)) do not show the same improvement, but instead sees a translation of the curve to lower ring loss so that the maximum $C_{\PC}$ now coincides with the larger $C_{\STM}$ values. Typical RC models show a trade-off between fading memory and nonlinearity \cite{Dambre2012}, however this trade-off is not as severe in the double-delay-line active-ring RC. $C_{\STM}$ and $C_{\PC}$ reach optimal values of $5.45 \pm 0.11$ and $2.13 \pm 0.06$ respectively for the same set of system parameters. \par

\section{Conclusion} \label{SectionV}

In conclusion, we have shown experimentally the physical implementation of the time-delay reservoir computing model using the spin-wave delay-line active-ring resonator operated with large values of delay time. We have shown that this system naturally satisfies the required reservoir properties of fading memory and nonlinearity due to the dynamics of the spin-wave delay line. By incorporating the reference line into the design of the RC, the system can exploit both the amplitude (nonlinear damping) and phase (nonlinear phase shift) nonlinearities of spin waves, which was shown to significantly improve its nonlinear computing capacity without drastic diminishing of the fading memory capacity. We evaluated performance using two benchmark tests and showed improved performance with regards to our previous RC concept and comparable to competing spintronic RC implementations. From an engineering standpoint, this system is very simple. Little pre-processing is required in order to inject data into the active ring and the linear readout can be performed directly on the microwave diode voltage, requiring no post-processing. Finally, we extended the delay time of the active ring by inserting a second delay line configured for the BVSW resulting in a partial compensation of the dispersive pulse broadening, improvement of the linear memory of the system and much better trade-off between linear memory and nonlinearity. \par

\section*{Acknowledgments}
The work of S. Watt was supported by the Australian Government Research Training Program.

\printbibliography[title={References}]

\end{document}